\documentclass[10pt]{article}
\usepackage{amsfonts}
\usepackage{mathrsfs}
\usepackage{amsmath}	
\usepackage{cite}
\usepackage{graphicx}
\usepackage{rotating}
\usepackage{hyperref}
\usepackage{verbatim}
\usepackage[all]{xy}

\csname @addtoreset\endcsname{equation}{section}
\newcommand{\beq}{\begin{equation}}
\newcommand{\eeq}{\end{equation}}
\newcommand{\bea}{\begin{eqnarray}}
\newcommand{\eea}{\end{eqnarray}}
\newcommand{\ba}{\begin{array}}
\newcommand{\ea}{\end{array}}
\newcommand{\bit}{\begin{itemize}}
\newcommand{\eit}{\end{itemize}}
\newcommand{\nn}{\nonumber}

\newcommand{\mezzo}{\frac{1}{2}}
\newcommand{\complesso}{{\ \hbox{{\rm I}\kern-.6em\hbox{\bf C}}}}
\newcommand{\reale}{{\hbox{{\rm I}\kern-.2em\hbox{\rm R}}}}
\newcommand{\uno}{ \,  \raisebox{+0.14em}{{\hbox{{\rm \scriptsize ]}} \raisebox{-0.2em}{\kern-.8em\hbox{1}}}} \, }  

\newcommand{\p}{\partial}

\renewcommand{\a}{\alpha}
\renewcommand{\b}{\beta}
\newcommand{\g}{\gamma}

\newcommand{\G}{\Gamma}
\renewcommand{\d}{\delta}
\newcommand{\D}{\Delta}

\renewcommand{\k}{\kappa}
\renewcommand{\l}{\lambda}
\renewcommand{\L}{\Lambda}

\newcommand{\m}{\mu}

\newcommand{\n}{\nu}
\renewcommand{\r}{\rho}
\newcommand{\s}{\sigma}

\newcommand{\z}{\zeta}

\newcommand{\om}{\omega}

\begin{document}


\begin{titlepage}
\begin{flushright}
CECS-PHY-15/05
\end{flushright}
\vspace{2.5cm}
\begin{center}
\renewcommand{\thefootnote}{\fnsymbol{footnote}}
{\huge \bf Microscopic Entropy of the Magnetised}
\vskip 5mm
{\huge \bf Extremal Reissner-Nordstrom Black Hole}
\vskip 30mm
{\large {Marco Astorino\footnote{marco.astorino@gmail.com}}}\\
\renewcommand{\thefootnote}{\arabic{footnote}}
\setcounter{footnote}{0}
\vskip 10mm
{\small \textit{
Centro de Estudios Cient\'{\i}ficos (CECs), Valdivia,\\ 
Chile\\}
}
\end{center}
\vspace{5.5 cm}
\begin{center}
{\bf Abstract}
\end{center}
{The extremal Reissner-Nordstr\"om black hole embedded in a Melvin-like magnetic universe is studied in the framework of the Kerr/CFT correspondence. The near horizon geometry can be written as a  warped and twisted product of $AdS_2 \times S^2$, also in presence of an axial external magnetic field that deforms the black hole. The central charge of the Virasoro algebra can be extracted from the asymptotic symmetries. Using the Cardy formula for the microscopic statistical entropy of the dual two-dimensional CFT, the Bekenstein-Hawking entropy, for this charged and magnetised black hole, is reproduced.}
\end{titlepage}








\section{Introduction}

In recent years progress has been made to describe the black hole entropy from a microscopical point of view. Even though, at the moment,  we do not possess a full consistent quantum theory of gravity, some speculations can be done in order to understand why, at the mascroscopical level, the entropy of a black hole is a quarter of its event horizon area. These results are based on some universal properties that any consistent theory of quantum gravity is expected to have, in the same spirit of the Boltzmann's statistical theory of gas. In fact, the universal laws of thermodynamics were derived, from statistical arguments, much before the discovery of quantum mechanics, which later brought the detailed microscopic atomic picture. \\
Although the tools used to describe the black hole microstates are inspired by holography and the AdS/CFT conjecture \cite{maldacena}, they basically rely on the symmetries of the diffeomorphism group of the classical spacetime and their relation to two-dimensional conformal field theory (CFT) \cite{brown-h}. Actually, it is not a coincidence that many features of a two-dimensional CFT are universal, i.e. they do not depend on the details of a given specific CFT. \\ 
The most relevant advance in this field concerns the inclusion in this framework of black holes of astrophysical interest such as the Kerr one \cite{strom08}, \cite{strom-castro} (for reviews see \cite{cargese}, \cite{compere-review}). It would be interesting to push forward the insight of the Kerr/CFT correspondence by considering a generalisation of the Kerr solution, that includes the presence of an axial external magnetic field. In fact, at the center of galaxies, where black holes are, most likely, thought to reside, an intense axial magnetic field is measured \cite{nature}. It is thought to be generated by matter falling, in the accretion disk, into the black hole. There are some exact analytical solutions that can model axisymmetric external electromagnetic field around a black hole. They are  generated by Ernst's solution generating technique \cite{ernst2}, \cite{ernst-magnetic}, by means of a Harrison transformation, and are interpreted as black holes embedded in a Melvin-like magnetic universe.\\
In this paper we will focus on the simple case of the magnetised Reissner-Nordstr\"om black hole, while in a following publication we will consider the most general case of the Kerr-Newman black hole immersed in an external magnetic universe\cite{magn-kerr-cft}. The reason to consider here the magnetised Reissner-Nordstr\"om solution is due to the fact that it is the simpler magnetised regular black hole which admits an extremal configuration, where the analysis, in terms of the Kerr/CFT approach, is clearer\footnote{Actually some tools of the Kerr/CFT correspondence can be applied also outside the extremality. They are based on the emergence of the conformal invariance in the wave equation of a probe field \cite{strom-castro},\cite{klemm}, on the fixed background geometry of the black hole.}. Indeed extremal black holes can be considered as a ground state from a thermodynamic point of view, because their Hawking temperature is null. Maybe this is the reason why the Kerr/CFT analysis is simpler  in that case. On the other hand non-extremal black holes, where their conserved charges (mass, angular momentum and electromagnetic charge) are unconstrained, can be considered as excited states. In these other cases the conformal symmetry of the near horizon geometry is not explicit.  \\
In fact the near horizon conformal symmetry is one of the fundamental ingredients of the Kerr/CFT method. This approach is of particular relevance in the context of magnetised black holes, because of their non trivial asymptotic, which is not asymptotically flat nor (A)dS. For this reason the thermodynamic properties of these spacetimes are not easy to treat and they are still not completely clear \cite{gibbons2}, \cite{booth}. Thus the Kerr/CFT, focusing only on near horizon properties, can bring a novel point of view to understand these potential interesting astrophysical objects\footnote{Although intense axial magnetic fields are measured at the center of galaxies, where massive black hole live, they may display different astrophysical characteristics with respect the Melvin-like magnetic fields we are here considering. Nonetheless these Ernst magnetised black holes are significant since they are the only analytical exact electrovacuum solution describing black holes in magnetic field and  may provide a good effective astronomical approximation, in some regime. }.  \\  
This paper is organised as follows. In section \ref{review} we review the basic features of the magnetised Reissner-Nordstr\"om black hole that will be used in rest of the paper. In section \ref{NHEMRN} the near horizon geometry of the extremal magnetised Reissner-Nordstr\"om solution is analysed. In section \ref{micro-entropy} the symmetries of the near horizon geometry are exploited to count the microscopical degrees of freedom that contribute to the entropy of the magnetised charged black hole.\\

\section{Magnetised Reissner-Nordstr\"om Black Hole Review}
\label{review}

We are interested in working in the standard framework of four-dimensional Einstein general relativity coupled with Maxwell electromagnetism, whose equations of motion\footnote{Henceforward  the Newton constant $G_N$ and the electromagnetic vacuum permeability $\m_0$ will be set to 1 for simplicity, without loss of generality. The Faraday tensor $F_{\m\n}$ is defined, as usual, from the electromagnetic vector potential $A_\m$ as $F_{\m\n} = \p_\m A_\n-\p_n A_\m $.}
\bea  \label{field-eq}
                        &&   \textrm{R}_{\m\n} -   \frac{\textrm{R}}{2}  g_{\m\n} = \frac{2G_N}{\m_0} \left( F_{\m\r}F_\n^{\ \r} - \frac{1}{4} g_{\m\n} F_{\r\s} F^{\r\s} \right)  \quad ,       \\
           \label{field-eq2}             &&   \partial_\m \big( \sqrt{-g} F^{\m\n} \big) = 0   \quad ,
\eea
stem from the Einstein-Maxwell action
\beq \label{action}
             I[g_{\m\n}, A_\m] =  -\frac{1}{16 \pi G_N}  \int_\mathcal{M} d^4x  \sqrt{-g} \left( \textrm{R} - \frac{G_N}{\m_0}F_{\m\n} F^{\m\n} \right)    \ \  .
\eeq
In this paper we will consider the magnetised Reissner-Nordstr\"om metric (MRN), which is a regular axisymmetric, stationary electrovacuum solution for the e.o.m. (\ref{field-eq}) - (\ref{field-eq2}). It describes an electrically\footnote{In this paper we will not consider a possible dyonic generalisation of the magnetised Reissner-Nordstr\"om spacetime which is additionally endowed with an intrinsic magnetic monopole black hole charge. This is because intrinsically magnetic charged black holes in external magnetic fields produces conical singularities that can not be removed, unless introducing a fine tuning with another physical parameter such as acceleration \cite{ernst-remove}, \cite{rot-pair-creation}.} charged black hole embedded in an axial axisymmetric magnetic field. MRN is not asymptotically flat, nor (A)dS. In fact, for large values of the radial coordinate $\tilde{r}$, the metric converges to the Melvin universe\footnote{Note that the electromagnetic field coincides with that of Melvin only up to a electromagnetic duality transformation \cite{gibbons2}.}. The Melvin magnetic universe is a static cylindrical symmetric spacetime containing an axial magnetic field. It represents a
universe containing a parallel bundle of electromagnetic flux held together by its own gravitational field. \\ 
The MRN solution can hardly be obtained by direct integration of the e.o.m. (\ref{field-eq}).  It was generated by Ernst \cite{ernst-magnetic}, through a Harrison transformation, from the standard static spherically symmetric Reissner-Nordstr\"om (RN) black hole. The metric and electromagnetic potential are given respectively by
\beq \label{MRN}
         ds^2  =   -\frac{f}{|\L|^2}(\D_\varphi d\tilde{\varphi}- \omega d{\tilde{t}})^2 + \frac{|\L| ^2}{f} \left[ \r^2 d{\tilde{t}}^2 - e^{-2\g} \left( \frac{d\tilde{r}^2}{\D_r} + \frac{dx^2}{\D_x} \right) \right]    \quad ,
\eeq
\beq \label{MA}
          A = A_{\tilde{t}} d{\tilde{t}} +A_{\tilde{\varphi}} d\tilde{\varphi} \quad ,
\eeq
with
\bea
        &&     \D_r(\tilde{r})  :=  \tilde{r}^2 + q^2-2m\tilde{r}         \quad  , \qquad            \D_x(x)  :=  1-x^2     \qquad ,  \qquad  \D_\varphi :=1+\frac{3}{2} q^2 B^2  + \frac{q^4 B^4}{16} \  , \quad  \\
         &&       f(\tilde{r},x):=-\tilde{r}^2\D_x     \qquad    \qquad   ,    \qquad             \rho^2 (\tilde{r},x)  := \D_r \D_x      \   \quad  , \qquad          e^{-2\g}(\tilde{r},x) := \tilde{r}^4 \D_x          \quad ,           \\
         &&  \om(\tilde{r},x):=  \frac{qB}{2\tilde{r}} \left\{B^2 \left[ \tilde{r}^2 + x^2 (\tilde{r}^2+q^2-2m\tilde{r} )  \right] -4 \right\}   \quad \  ,       \\
         &&   \L(\tilde{r},x) := 1 - i q B x + {B^2 \over 4}  \left[ \tilde{r}^2 (1-x^2) + q^2 x^2 \right]    \quad \qquad \label{Lambda},        \\
         &&  A_{\tilde{\varphi}}(\tilde{r},x) := A_{\tilde{\varphi}_0} + \frac{B \D_\varphi}{8 |\L|^2} \left[ 12 q^2 x^2 + B^2 q^4 x4 - B^2 \tilde{r}^4 (1-x^2)^2 + 2 \tilde{r}^2 (1-x^2) (2+ q^2 B^2 x^2)  \right]   \  , \ \    \\
         &&  A_{\tilde{t}} (\tilde{r},x):=  A_{\tilde{t}_0} + \om \left(\frac{3}{2B} - \frac{A_{\tilde{\varphi}} - A_{\tilde{\varphi}_0} }{\D_\varphi}\right) + \frac{2q}{\tilde{r}}  \quad ,     \label{At}
\eea
where the $m, q, B$ constants parametrise the mass, the intrinsic charge of the black hole and the external magnetic field respectively. While the constant $\D_\varphi$ is explicitly added to avoid conical singularities on the symmetry axis $(\tilde{r}=0, x=\pm 1)$ without changing the usual range of the azimuthal angular coordinate $\tilde{\varphi} \in [0,2\pi)$. The other angular coordinate is $x = \cos \theta   \in [-1,1)$. 
When $B=0$  the solution (\ref{MRN})-(\ref{MA}) exactly recover the RN black hole, in that case $m$ and $q$ coincides with the mass and electric charge of the black hole. While for vanishing $m$ and $q$ we retrieve the Melvin magnetic universe.  The constants  $A_{\tilde{\varphi}_0}$ and $A_{\tilde{t}_0}=0$ are not relevant for the equations of motion but they appear in conserved quantities. In order to have a regular electromagnetic field on the rotation axis, $A_{\tilde{\varphi}_0}$ can be gauge fixed such that  $A_{\tilde{\varphi}}=0$ at $x=\pm1$:
\beq \label{gauge-A}
            \bar{A}_{\tilde{\varphi}_0} = -  \frac{1}{8} B q^2 (12+q^2 B^2)
\eeq
For simplicity $A_{\tilde{t}_0}=0$ will be fixed to 0, coherently with the unmagnetised case.
As the standard RN black hole, the metric (\ref{MRN})-(\ref{Lambda}) possesses an inner $\tilde{r}_-$ and an outer (event) horizon $\tilde{r}_+$  positioned at
\beq
                    \tilde{r}_\pm = m \pm \sqrt{m^2-q^2} \qquad .
\eeq 
In this paper we will mostly focus on the extremal limit of the  magnetised Reissner-Nordstr\"om solution (\ref{MRN})-(\ref{At}), (henceforth EMRN) which can be obtained, as in the case without the external magnetic field, in the limit $q \rightarrow m$, therefore the inner and outer horizon coincide at a double degenerate event horizon $\tilde{r}_e=m$. \\
The presence of the axial external magnetic field modifies the geometry of the black hole horizon with respect to the spherical RN case, which for the MRN become oblate along the direction of the magnetic flux. In fact inspecting the black hole area
\beq \label{area}
     \mathcal{A} = \int_0^{2\pi} d\tilde{\varphi} \int_{-1}^1 dx  \sqrt{g_{\tilde{\varphi}\tilde{\varphi}}g_{xx}} \Big|_{\tilde{r}=\tilde{r}_+} = 4 \pi \D_\varphi \tilde{r}^2_+
\eeq
and the equatorial $C_{eq}$ circumference of the event horizon (for constant ${\tilde{t}}$ and $x=0$)
\beq
      C_{eq} =  \int_0^{2\pi} \sqrt{g_{\tilde{\varphi}\tilde{\varphi}}} \ d\tilde{\varphi} \Big|_{\tilde{r}=\tilde{r}_+} = \frac{2 \pi \D_\varphi \tilde{r}_+}{1+B^2 \tilde{r}^2/4} \quad ,
\eeq
we can appreciate the stretching effect of the external magnetic field. The fact that the MRN black hole area is different with respect to the $B=0$ case (i.e. RN), suggests that, in general, the Harrison transformation modifies the entropy of the black hole. To be more precise the area is modified by the factor $\D_\varphi$ introduced to avoid conical singularities after the Harrison transformation. Only in some special cases, as the Schwarzschild-Melvin solution (i.e. setting $q=0$ in (\ref{MRN})-(\ref{At})), where the Harrison transformation does not affect the angular momentum, the black hole entropy remains unchanged\footnote{In \cite{klemm} this point was left open.}.  \\
Moreover the interaction of the external magnetic field with the intrinsic electric charge of the Reissner-Nordstr\"om black hole generates a Lorentz force, which makes the spacetime stationary spinning, with angular momentum given in (\ref{ang-mom}). This fact makes the MRN black hole treatment, in the framework of Kerr/CFT correspondence, closer to the Kerr-Newman case instead of the standard RN.\\
The conserved charges were discussed, only recently, for magnetised black hole in \cite{gibbons2} and \cite{booth}, unfortunately there is not agreement on the value of the mass. We will use the approach of \cite{barnich}, which for angular momentum it is basically equivalent to the Komar integral
\bea \label{ang-mom}
       J &=& \frac{1}{16\pi} \lim_{\mathcal{S}_t\rightarrow \infty} \int_ {\mathcal{S}_t} \left[ \nabla^\a \xi^\b_{(\varphi)} + 2 F^{\a\b} \xi^\m_{(\varphi)} A_\m \right] dS_{\a\b}   \\
       &=& \left[  q A_{\tilde{\varphi}_0}  - \frac{1}{4} B q^3 \left( B A_{\tilde{\varphi}_0} -2 \right) - \frac{1}{2} B^3 q^5 - \frac{1}{32} B^5 q^7 \right]  \Big|_{ A_{\tilde{\varphi}_0}=\bar{A}_{\tilde{\varphi}_0}} = - B q^3 - \frac{1}{4} B^3 q^5  \ \ , \quad 
\eea
where $\xi^\m_{(\varphi)}$ represent the rotational Killing vector $\p_{\tilde{\varphi}}$ and where $dS_{\a\b}= - 2 n_{[\a}\s_{\b]} \sqrt{g_\mathcal{S}} \ d\tilde{\varphi} dx$, with $\sqrt{g_\mathcal{S}} = \sqrt{g_{xx} g_{\tilde{\varphi} \tilde{\varphi}}}$, is the infinitesimal volume element of the two-dimensional surface $\mathcal{S}_t$ (obtained for fixed time and radial coordinates) surrounding the black hole horizon we are integrating over. $n_\m$ and $\s_\n$ are two orthonormal vectors, timelike and spacelike respectively, normal to the surface of integration $\mathcal{S}_t$. Note that only if the electromagnetic potential is property regularised by the gauge fixing (\ref{gauge-A}), the value of $J$ coincides with that of \cite{gibbons2} and \cite{booth}.  \\
The electric charge is given by
\beq
        \mathcal{Q} =\frac{1}{8\pi} \int_\mathcal{S} F^{\m\n} dS_{\m\n} = - \frac{1}{4\pi} \int_0^{2\pi} d\tilde{\varphi}  \int_{-1}^1 dx \ \sqrt{g_\mathcal{S}} \ n_\m \s_\n F^{\m\n}  =  q \left( 1 - \frac{1}{4} q^2 B^2 \right)   \quad ,
\eeq
while the magnetic charge is $P=\frac{1}{4\pi} \int_\mathcal{S} F$  is null  because the black hole we are considering does not have intrinsic magnetic monopoles\footnote{It is possible to build also a dyonic Reissner-Nordstr\"om black hole, with both electric and magnetic charge, embedded in the Melvin external magnetic field. But when the external field and the intrinsic charge of the black hole are of the same kind (magnetic for instance) the metric presents asymmetric conical singularities on the poles, that can not be removed just by an azimuthal angular rescaling. Eventually they can be removed by an additional parameter, such as the one provided by  acceleration, as done for instance in \cite{ernst-remove} or \cite{rot-pair-creation}.}, just an external magnetic field, whose amount of entering flux lines is equal to the outgoing ones.   \\
In the following sections will be useful to know the  electrostatic Coulomb potential, conjugated to the electric charge, $\Phi_e$ (and its extremal limit $\Phi^{ext}_e$) of the solution, which is defined on the horizon as
 \bea \label{Phi}
               \Phi_e := -\chi^\m A_\m\big|_{\tilde{r}=\tilde{r}_+}   &=&  - \frac{2 B \bar{A}_{\tilde{\varphi}_0} }{\D_\varphi q} \left[ \left( -1 + \frac{q^2 B^2}{4}\right) m + \left( 1+ \frac{q^2 B^2}{4}\right) \sqrt{m^2-q^2}\right] +\\ 
               &+& \frac{1}{q} \left[ \left( -1 + \frac{3 q^2 B^2}{4}\right) m + \left( 1+ \frac{3 q^2 B^2}{4}\right) \sqrt{m^2-q^2}\right]  \quad \qquad , \\
               \ \Phi_e^{ext} :=   \lim_{q\rightarrow m} \Phi_e  \ &=& \ 1 + \frac{2 B \bar{A}_{\tilde{\varphi}_0} }{\D_\varphi} - \frac{m^2 B^2}{4 \D_\varphi} (2 B \bar{A}_{\tilde{\varphi}_0} + 3 \D_\varphi)      \quad ,
 \eea
where the Killing vector $\chi:= \p_t + \Omega_J \p_\varphi$ is the tangent to the horizon's null generator\footnote{The vector $\chi$ remains a Killing vector even if rescaled by a constant factor $\chi \rightarrow\a \chi$. While in the Kerr geometry there is a natural way to set this normalisation, in presence of the magnetic field it is not clear how to fix this scaling \cite{booth}. Here we keep using the standard definition, that has the advantages of reproducing the known cases in the limit of vanishing external magnetic field.}. 
The chemical potential $\Omega_J$, conjugate to the angular momentum $J$, represents the angular velocity of the black hole horizon, it is given by
\beq \label{Omega}
                \Omega_J = -\frac{g_{\tilde{t} \tilde{\varphi}}}{g_{\tilde{\varphi}\tilde{\varphi}}} \ \Big|_{\tilde{r}=\tilde{r}_+} = - \frac{2 q B}{\D_\varphi \  \tilde{r}_+} + \frac{ q B^3}{2 \D_\varphi} \ \stackrel{q\rightarrow m}{\longrightarrow} \ \Omega_J^{ext} = \frac{B(m^2B^2-4)}{2 \D_\varphi} \quad . \quad
\eeq

\section{Near horizon extremal magnetised Reissner-Nordstr\"om}
\label{NHEMRN}

We want to analyse the region very near the extreme magnetised Reissner-Nordstr\"om event horizon $\tilde{r}_e$. In order to do so, following  \cite{Bardeen:1999px} and \cite{compere-review}, we define new dimensionless coordinates ($t,r,\varphi$) in this way:
\beq
            \tilde{r}(r) := r_e + \l r_0 r    \qquad , \qquad      \tilde{t}(t) := \frac{r_0}{\l} t     \qquad  ,    \qquad     \tilde{\varphi}(\varphi,t):=\varphi + \Omega_J^{ext} \ \frac{r_0}{\l} t  \quad  , \quad
\eeq
where $r_0$ is introduced to factor out the overall scale of the near-horizon geometry.
In  the presence of the electromagnetic potential $A_\m$, before taking the near-horizon limit, we need to perform the gauge transformation
\beq
          A_{\tilde{t}} \rightarrow A_{\tilde{t}} + \Phi_e  \quad .
\eeq 
The near horizon, extreme, magnetised, Reissner-Nordstr\"om geometry (NHEMRN) is defined as the limit of the EMRN for $\l \rightarrow 0$. 
Remarkably the NHEMRN geometry can be cast in the general form of the near-horizon geometry of standard extremal spinning black holes. This form posses a $SL(2,\mathbb{R}) \times U(1)$ isometry\cite{compere-review}, which is a warped and twisted product of $AdS_2 \times S^2$, given by
\beq \label{near-metric}
           ds^2 = \G(x) \left[ -r^2 dt^2 + \frac{dr^2}{r^2} + \a^2(x) \frac{dx^2}{1-x^2} + \g^2(x) \ \big( d\varphi + \k r dt \big)^2 \right] \quad ,
\eeq 
where in our case we set $r_0=m$, as in the unmagnetised RN case, and
\bea \label{fields1}
           \G(x) &=& m^2 \left[ (1+m^2 B^2/4)^2 + m^2 B^2 x^2  \right]     \qquad , \qquad \a(x)= 1    \qquad \qquad  \qquad \ \ \quad  , \qquad   \\
    \label{fields2}    \g(x)  &=& \frac{m^2 \D_\varphi \sqrt{1-x^2}}{\G(x)}       \qquad    \qquad  \qquad \qquad   \ \quad , \qquad \k = -\frac{mB(4+m^2B^2)}{2 \ \D_\varphi}    \qquad . \quad 
\eea
While the near horizon electromagnetic one form is described by
\beq \label{near-A}
               A = \ell(x) (d\varphi+\k r dt) - \frac{e}{\k} d\varphi \qquad ,
\eeq
where
\beq
           \ell(x) = -\frac{\D_\varphi r_0}{2 m^2 B} \frac{(4-m^2B^2)[(1+m^2B^2/4)^2-m^2B^2x^2]}{(4+m^2B^2)\left[(1+m^2B^2/4)^2+m^2B^2x^2\right]} \qquad ,  \qquad e = m \left(1+\frac{3}{4} m^2 B^2 \right) - \bar{A}_{\tilde{\varphi}_0} \k  \ \ .\ 
\eeq
This relevant aspect is, maybe,  not surprising in view of \cite{kunduri}, where it is shown that, for the theory under consideration,  governed by the action (\ref{action}), any extremal black hole near horizon geometry has the form (\ref{near-metric}).\\
The $SL(2,\mathbb{R}) \times U(1)$ isometry of the NHEMRN is infinitesimally generated by the following Killing vectors
\bea
              \z_{-1} = \p_t  \qquad &,& \qquad \z_0 = t \p_t - r \p_r \qquad \\
              \z_1 = \left( \frac{1}{2r^2} +\frac{t^2}{2} \right) \p_t - t r \ \p_r - \frac{\k}{r} \p_\varphi \qquad  &,& \qquad L_0 = \p_\varphi \qquad .          
\eea
From their commutation relations (we write only the not null commutators)
\bea
              [\z_0 , \z_{\pm}] = \pm \z_{\pm}   \qquad \quad , \quad \qquad [ \z_{-1} , \z_1 ] = \z_0 \qquad       
\eea 
we can infer that $\{\z_{-1}, \z_0, \z_{1}\}$ form the $SL(2,\mathbb{R}) \sim SO(2,1)$ algebra, while $L_0$ constitutes the $U(1)$ algebra.  The normalisation of the generators of infinitesimal isometries is chosen to simplify the commutation rules.\\
The fact that the external magnetic field is not breaking the black hole near horizon symmetry is the fundamental point in the microscopic analysis of the entropy, as will be done in section \ref{micro-entropy}.

\subsection{Boundary conditions}
\label{subsec-boundary}

Inspired by the AdS/CFT conjecture we suppose that the non-trivial thermodynamic properties of these extremal black holes manifest on the boundary of the near-horizon geometry, as done in the standard Kerr/CFT formulation \cite{strom08}, \cite{compere-review}. To implement this view we have to find the asymptotic symmetries (for large $r$) of the near horizon geometry to include a copy of the Virasoro algebra.  In \cite{stro-duals} the following generators are proposed
\bea 
        \z_\epsilon &=&  \epsilon(\varphi) \p_\varphi -r \epsilon'(\varphi)  \p_r \  +\  \textrm{subleading terms} \quad  , \quad  \label{zep}\\
        \xi_\epsilon &=& -\left[\ell(\theta)-\frac{e}{\k} \right]  \epsilon(\varphi ) \   + \ \textrm{subleading terms}  \quad . \quad    \label{xiep}
\eea  
They preserve the following fall-off boundary conditions \cite{compere-review} for the metric (\ref{near-metric})
\begin{align}
g_{tt}&= \mathcal{O}\left({r^2}\right),\qquad g_{t\varphi}= \k \Gamma(x)\gamma(x)^2 r + \mathcal{O}\left({1}\right),\nn\\
g_{tx} &= \mathcal{O}\left({1\over r}\right),\qquad g_{t r}= \mathcal{O} \left({1\over r^2}\right) ,\qquad g_{\varphi \varphi}= \mathcal{O}(1), \nn \\
g_{\varphi x}&= \mathcal{O}\left({1\over r}\right) ,\qquad g_{\varphi r}= \mathcal{O} \left({1\over r}\right) ,\qquad g_{x r}= \mathcal{O} \left({1\over r^2}\right) , \label{bound-cond} \\
g_{xx}&= \frac{\Gamma(x)\alpha(x)^2}{\D_x} + \mathcal{O} \left({1\over r}\right) ,\qquad 
g_{rr}= \frac{\Gamma(x)}{r^2}+\mathcal{O} \left({1\over r^3}\right),\nn
\end{align}
and for the electromagnetic one-form
\begin{align}
	A_{t}&= \mathcal{O}\left({r}\right) \quad ,\qquad A_{\varphi}= \ell(\theta) - \frac{e}{\k} + \mathcal{O}\left({\frac{1}{r}}\right),\nn\\
	A_{x} &= \mathcal{O}\left(1\right) \quad,\qquad A_{r}= \mathcal{O} \left({1\over r^2}\right) \quad .
\end{align}

In \cite{stro-duals} also a zero energy and electric charge excitation conditions  are imposed for the charges $ \d \mathcal{Q}_{\p_t} = 0 $ and $ \d \mathcal{Q} = 0 $, respectively.
Note that these fall-off condition are preserved by $\z_\epsilon$ and by $\{\z_{-1},\z_0\}$ but not by the whole $SL(2,\mathbb{R})$ symmetry of the near horizon geometry on the bulk. \\
For the NHEMRN we will borrow these boundary conditions (\ref{bound-cond}), originally proposed for the near horizon geometry of the extremal Kerr black hole.

\section{Microscopic entropy}
\label{micro-entropy}

The appearance of the the Virasoro asymptotic algebra motivates the conjecture that quantum gravity on the NHMERN region is dual to a two-dimensional CFT on its boundary. From the Dirac bracket between the charges associated with the asymptotic symmetry vectors (\ref{zep})-(\ref{xiep}) it is possible to find the central extension of the Virasoro algebra \cite{cargese},\cite{compere-review}.  As explained in \cite{compere-review}, for the general near horizon geometry described by (\ref{near-metric}) and governed by the action (\ref{action}), it is possible to prove that matter does not contribute directly to the value of the central charge, but only influence it through the functions $\G(x),\g(x),\a(x)$ and $\k$.\\
In practice the central charge can be simply computed as the $m^3$ factor in the following equation
\beq
                c_J = 12 i \lim_{r \rightarrow \infty} \mathcal{Q}_{L_m} ^{\mathrm{Einstein}} [\mathcal{L}_{L_{-m}} \bar{g}; \bar{g}] \Big|_{m^3} \quad ,
\eeq
where $\mathcal{L}_{L_{-m}} \bar{g}$ is the Lie derivative of the metric along $L_{-m}$ and the fundamental charge formula for Einstein gravity \cite{35} is given by
\beq
\mathcal{Q}_{L_{m}}^{\mathrm{Einstein}} [h ; \bar{g}] = \frac{1}{8 \pi G_N}\int_S dS_{\mu\nu}\Big( \xi^\nu \nabla^\mu h +\xi^\mu \nabla_\sigma h^{\sigma\nu}+\xi_\sigma \nabla^\nu h^{\sigma\mu} +\frac{1}{2}h \nabla^\nu \xi^\mu +\frac{1}{2}h^{\mu\sigma}\nabla_\sigma \xi^\nu+\frac{1}{2}h^{\nu\sigma}\nabla^\mu \xi_\sigma\Big)\ . \nn
\eeq
Here  $h:=\bar{g}^{\m\n} h_{\m\n}$, while $\mathcal{S}$ and $dS_{\m\n}$ are defined as in section \ref{review}. $\mathcal{Q}_{L_{m}}^{\mathrm{Einstein}} [h ; \bar{g}]$ represents the charge, associated with the Killing vector $\xi^\m$, of the linearised metric $h_{\m\n}$ around the background spacetime $\bar{g}_{\m\n}$. Using the near horizon field behaviour expressed in eqs. (\ref{near-metric}), (\ref{near-A}) and the its asymptotic symmetry generator (\ref{zep}) one obtains
\beq \label{cc}
         c_J = \frac{3\ \k}{G_N \hbar} \int_{-1}^1 \frac{dx}{\sqrt{1-x^2}} \ \G(x) \a(x) \g(x) \quad .
\eeq 
In particular substituting in (\ref{cc}) the EMRN fields (\ref{fields1})-(\ref{fields2}) and performing the integration we obtain the actual value of the central charge for the extremal magnetised Reissner-Nordstr\"om black hole\footnote{We will use units where $G_N=1, \hbar=1$.}
\beq
          c_J = -3 m^3 B (4+m^2 B^2)    \quad .
\eeq
Exploiting the assumption that the near horizon geometry of the extremal Reissner-Nordstr\"om black hole embedded in a Melvin-like external magnetic field is described by the left sector of a two-dimensional CFT, we can apply the Cardy formula
\beq \label{cardy}
             \mathcal{S}_{CFT} = \frac{\pi^2}{3} (c_L T_L+c_R T_R)
\eeq
to compute the microscopic entropy of the dual systems\footnote{Fortunately the details of the CFT on the boundary are not necessary to describe the thermodynamic behaviour of the black hole in the bulk.}. Since we are considering only the rotational excitations along $\p_\phi$, corresponding to the left sector, the right modes will be considered frozen, thus their temperature $T_R$ will vanish.  On the other hand, as the left temperature of the CFT, we associate the Frolov-Thorne temperature $T_\varphi$ of the black hole, which near the horizon is well defined. We are using this temperature instead of the Hawking one because this latter is zero at extremality, but still quantum states just outside the horizon are not pure states when one defines the vacuum using the generator of the horizon \cite{strom08}, \cite{stro-duals}. In fact the Frolov-Thorne temperature, in the context of quantum field theory on curved background,  takes into account these rotational excitation degrees of freedom generalising the Hartle-Hawking vacuum, originally built for the Schwarzschild black hole. The extremal Frolov-Thorne temperature can be deduced by kinematical quantities, thus it is encoded in the metric and the matter fields, without directly using the equations of motion. In general it can be obtained as the following limit to extremality, that is $\tilde{r}_+\rightarrow \tilde{r}_e$ (which in our case it means $q \rightarrow m$)
\beq \label{Tphi}
            T_\varphi :=   \lim_{\tilde{r}_+\rightarrow \tilde{r}_e}  \frac{T_H}{\Omega_J^{ext}-\Omega_J}  = \frac{-\D_\varphi}{\pi m B (4 + m^2 B^2)} = \frac{1}{2 \pi \k }   \qquad ,
\eeq
where the Hawking temperature  $T_H$ is given in terms of the surface gravity $k_s$ as follow:
\beq 
            T_H  :=  \frac{\hbar \ k_s}{2 \pi} =  \frac{\hbar}{2 \pi} \sqrt{-\mezzo \nabla_\m \chi_\n \nabla^\m \chi^\n} = \frac{\hbar}{2 \pi}  \ \frac{\tilde{r}_+ - \tilde{r}_-}{2 r_+^2} \qquad .
\eeq
It can be easily noted as for extreme black holes, also in the presence of the external magnetic field, the surface gravity and Hawking temperature are null because inner and outer horizons concide: $\tilde{r}_+=\tilde{r}_-$. \\
The surface gravity remains exactly the same of the Reissner-Nordstr\"om metric and the Frolov-Thorne temperature is of the same form (but different value) of the standard non magnetised black holes, $T_\varphi =(2 \pi \k)^{-1}$. Analogously we can compute the electric chemical potential $T_e $ in presence of an external magnetic field 
\beq \label{Te}
   T_e :=   \lim_{\tilde{r}_+\rightarrow \tilde{r}_e}  \frac{T_H}{\Phi_e^{ext}-\Phi_e} = \frac{1}{2 \pi \left[ m \left(1 + \frac{3}{4} m^2 B^2 \right) -  \bar{A}_{\tilde{\varphi}_0} \k \right]} \qquad ,
\eeq
where the electrostatic potentials $\Phi_e$ and $\Phi_e^{ext}$  were defined in (\ref{Phi}). It is a non trivial fact that $T_e$ in (\ref{Te}), also for the extremal magnetised Reissner-Nordstr\"om black hole, can be simply expressed as
\beq \label{Tef}
         T_e = {1\over 2 \pi e} \quad .
\eeq 
This form of $T_e$ is exactly the same as the whole (AdS)-Kerr-Newman spacetime family and, as it is conjectured  in \cite{compere-review}, we have just shown that it can be suitable also for a more general class of black holes.\\
Then using as the left temperature $T_\varphi$ (\ref{Tphi}) in the Cardy formula (\ref{cardy}) we get the microscopic entropy of the EMRN black hole
\beq \label{entropy}
           \mathcal{S_{CFT}}=\frac{\pi^2}{3} c_J T_\varphi = \pi m^2 \D_\varphi = \frac{1}{4} \mathcal{A}^{ext}
\eeq
Note that the entropy of the dual CFT system coincides with a quarter of the extremal black hole area (\ref{area}), as happens for Bekenstein-Hawking entropy of non-magnetised black holes.\\
It is interesting to observe that, although the electromagnetic field is highly non trivial, the entropy is matched precisely only with the contribution of the gravitational central charge. This fact is in complete analogy with the (extreme) AdS-Kerr-Newman case where the contribution to the central charge of the gauge field is null.\\
Note also that when the external magnetic field vanishes we recover the usual result for the entropy of the extremal Reissner-Nordstr\"om black hole. Of course when $B=0$ some quantities we have used to get the entropy can not be defined, such as $T_\varphi$ because the black hole is static. However, as explained in \cite{compere-review} for the unmagnetised extremal RN spacetime, it is still possible to use another CFT description based on the five-dimensional Kaluza-Klein uplift of the electromagnetic gauge field along a compact direction generated by the Killing vector $\p_\psi$, which can be considered an angle of lenght $2\pi R_\psi$. Thus it is possible to define a chemical potential associated with the $\p_\psi$ direction. Because the extra-dimensional angular coordinate has a period of $ 2 \pi R_\psi $, the extremal Frolov-Thorne temperature is expressed in units of  $R_\psi$
\beq  \label{Tpsi}
             T_\psi = T_e R_\psi = \frac{R_\psi}{2 \pi e}    \quad .
\eeq 
Thinking the electrically charged particles immersed in a thermal bath with temperature $T_\psi$, it is possible to identify the the left sector of the dual CFT with a density matrix at temperature $T_\psi$. Therefore, assuming again that there are not right excitations at extremality, instead of associate $T_L$ with the left-moving temperature $T_\varphi$,  we can assign
\beq  \label{assi}
                              T_L = T_\psi  \qquad , \qquad T_R  = 0  \quad .
\eeq  
Also in this electromagnetic picture is possible to compute the central charge $c_Q$, by uplifting the solution in higher dimensions, as done for the Kerr-Newman black hole \cite{compere-review}, \cite{204}
\beq \label{ccQ}
          c_Q =  \frac{3\ e}{R_\psi} \int_{-1}^1 \frac{dx}{\sqrt{1-x^2}} \ \G(x) \a(x) \g(x)    \quad .
\eeq
If we substitute $c_Q$ as the left central charge into the Cardy formula (\ref{cardy}), thanks to the equations (\ref{Tpsi})-(\ref{ccQ}) we obtain again the entropy for the EMRN black hole $\mathcal{S_{CFT}}= \pi m^2 \D_\varphi = \mathcal{A}^{ext}/4$, exactly as in (\ref{entropy}). Note that in this other dual picture not only the resulting entropy is converging to the unmagnetised case when $B=0$, but also all the intermediate steps are compatible with the vanishing of the external magnetic field. In fact this is the usual procedure followed to find the microscopic  entropy of the (not magnetised, thus static) extremal RN black hole \cite{stro-duals}. \\
It would be interesting to extend these results outside the extremal limit. Unfortunately the tools of Kerr/CFT correspondence are less effective when the black hole is not extremal, because a reduction of the near horizon symmetry. Nevertheless when the external magnetic field is not present some speculation outside extremality can be done. They are based on the symmetry of the wave equation of a (not self-interacting) probe scalar field  in the near-region\footnote{The near region approximation does not coincides with the near horizon one.}  of the black hole, which is considered a fixed background. Then introducing some extra assumption on the value of the central charges, which are both switched on, in the not extremal case, it is possible to reproduce the entropy via the Cardy formula (\ref{cardy}).  \\
In presence of the external magnetic field the wave equation is not separable any more, except in the case of weak field, that is when only the linear contribution in $B$ is considered. Thus the usual non-extremal analysis can not be accomplished.

\section{Summary and Comments}

In this paper the tools provided by the Kerr/CFT correspondence are used to study extremal Reissner-Nordstr\"om black hole embedded in a Melvin-like magnetic universe. We have found that the symmetries of the near horizon geometry are not broken by the presence of the external magnetic field. Therefore  near the horizon the metric can be written as a warped and twisted product of $AdS_2 \times S^2$  endowed with the $SL(2,\mathbb{R}) \times U(1)$ isometry. This opens to the possibility of describing some features of the gravitational theory through the duality with a conformal field theory in two-dimension.  Borrowing the standard boundary conditions used in the Kerr(-Newman)/CFT correspondence we were able to compute how the central charge of the Virasoro algebra, which generate the asymptotic symmetries, is deformed by the presence of the external magnetic field.  Also the Frolov-Thorne temperature, thought preserving the standard form, is affected by the presence of the magnetic field. Using the Cardy formula for the microscopic statistical entropy of the dual two-dimensional CFT the Bekenstein-Hawking entropy of the black hole is exactly reproduced.  \\
In this picture the CFT is related to the rotational excitations along the $\p_\varphi$ Killing vector, created by the Lorentz interaction between the intrinsic electric charge of the black hole and the external magnetic field\footnote{Note that this procedure is not viable in the standard non-magnetised extremal RN case, nevertheless, for the magnetised RN we are considering, the entropy remains well defined also in the limit of vanishing external magnetic field $ B \rightarrow 0 $.}. 
But, in presence of electrically charged black holes, another dual CFT description is available. The latter is related to the $U(1)$ symmetry of the electromagnetic gauge potential.  Also in this case it is possible to reproduce the classical entropy by the CFT methods. A non trivial result is that also in this magnetised case the left temperature of the dual CFT, related to the electromagnetic chemical potential on the bulk gravity side, remains of the same form (\ref{Tef}) of the AdS-Kerr-Newman family of black holes, although influenced by the external magnetic field.\\  
Therefore the Kerr/CFT correspondence has been shown to work well also in the presence of axial external magnetic fields, which deform the asymptotics in a non trivial way. This is an encouraging step towards a further generalisation to magnetised black holes of the Kerr-Newman family.  Work in this direction is in progress \cite{magn-kerr-cft}. \\
Unfortunately the extension to non extremal magnetised metrics is not so straightforward  because the usual tools developed in Kerr/CFT correspondence are not working, due to a reduction of the symmetry outside extremality.

\section*{Acknowledgements}
\small I would like to thank  Cristian Erices, Timothy Taves and in particular Cedric Troessaert for fruitful discussions. \\ 
The Centro de Estudios Cient\'{\i}ficos (CECs) is funded by the Chilean Government through the Centers of Excellence Base Financing Program of Conicyt. 
\normalsize


\end{document}